\documentclass{article}

 \usepackage[dblblindworkshop, final]{neurips_2025} \workshoptitle{ResponsibleFM}
\usepackage[utf8]{inputenc} % allow utf-8 input
\usepackage[T1]{fontenc}    % use 8-bit T1 fonts
\usepackage{hyperref}       % hyperlinks
\usepackage{url}            % simple URL typesetting
\usepackage{booktabs}       % professional-quality tables
\usepackage{amsfonts}       % blackboard math symbols
\usepackage{nicefrac}       % compact symbols for 1/2, etc.
\usepackage{microtype}      % microtypography
\usepackage{xcolor}         % colors
\usepackage{adjustbox}

\title{Rethinking AI Evaluation through TEACH-AI: A Human-Centered Benchmark and Toolkit for Evaluating AI Assistants in Education}

\author{%
  Shi Ding \\
  Expressive Machinery Lab \\
  Georgia Institute of Technology \\
  Atlanta, GA 30332 \\
  \texttt{sding84@gatech.edu} \\
  \And
  Brian Magerko \\
  Expressive Machinery Lab \\
  Georgia Institute of Technology \\
  Atlanta, GA 30332 \\
  \texttt{magerko@gatech.edu} \\
}

\usepackage{booktabs}
\usepackage{tabularx}
\usepackage{array}

\begin{document}

\maketitle

\begin{abstract}

As generative artificial intelligence (AI) continues to transform education, most existing AI evaluations rely primarily on technical performance metrics such as accuracy or task efficiency while overlooking human identity, learner agency, contextual learning processes, and ethical considerations. In this paper, we present \textit{TEACH-AI (Trustworthy and Effective AI Classroom Heuristics)}—a domain-independent, pedagogically grounded, and stakeholder-aligned benchmark framework with measurable indicators and a practical toolkit for guiding the design, development, and evaluation of generative AI systems in educational contexts. Built on an extensive literature review and synthesis, the ten-component assessment framework and toolkit checklist provide a foundation for scalable, value-aligned AI evaluation in education. TEACH-AI rethinks “evaluation” through sociotechnical, educational, theoretical, and applied lenses, engaging designers, developers, researchers, and policymakers across AI and education. Our work invites the community to reconsider what constructs “effective” AI in education and to design model evaluation approaches that promote co-creation, inclusivity, and long-term human, social, and educational impact.

\end{abstract}


\section{Introduction and Related Work}

% Problems 
As generative AI systems increasingly become intelligent assistant in learning environments, they challenge the traditional roles of teachers, tools, and humans \cite{shetye2024evaluation, thomas2024using}. AI education interventions are often designed for and evaluated based on the efficacy of the AI technology in terms of its behavior, sensing capabilities, and reasoning \cite{sperrle2021survey, vanlehn2006behavior} centered on agent-human interactions. Rarely do these works involve the broader learning context of their designs and evaluations \cite{resnick2024generative, shaffer1999thick, friedman2002value, mcklin2018authenticity}. Therefore, rather than benchmarking against the status quo or competing models, this article attempts to enable researchers to evaluate how well their AI-based interventions work with a multi-faceted framework and a practical toolkit that captures both the top-down and bottom-up factors related to human success with generative AI tools. We aim to investigate our research question: \textit{What criteria define effective, value-aligned human–AI collaboration in educational settings, and how might these criteria guide the development of the TEACH-AI benchmark and practical toolkit for evaluating generative AI assisting tools?}

% Related Work: show landscape + limits + %Thematic Gap: Organize lit review + show unmet needs

\subsection{Generative AI in Educational Contexts}

Recent research highlights generative AI (GenAI)’s potential to support multiple areas of education, including: a) educational administration by reducing teachers' workload (e.g., auto-grading and instructional content generation)\cite{ali2021social}; b) personalized learning\cite{vincent2020trustworthy}, c) digital literacy development \cite{eshet2004digital, bundy2017preparing}, and d) with growing interest in its role in developing higher order thinking skills\cite{baer1998case, amabile2018creativity, grover2013computational}. AI-supported learning aligns with theories like Zone of Proximal Development(ZPD)\cite{vygotsky1978mind} and Constructionism \cite{papert1991situating}, which emphasize contextualized support and human agency. However, GenAI’s effectiveness in K–12 and higher education remains underexplored due to ongoing concerns about over-reliance on Generative AI, socio-emotional and creative limitations in AI-generated feedback, and broader ethical risks including bias, factual inaccuracies, trust, equity, and accountability \cite{resnick2024generative, alasadi2023generative}. 

Addressing these issues requires design and evaluation approaches aligned with human identities, values, and community norms \cite{vincent2020trustworthy, ash2020reimagining, friedman2019value}. In response, we propose a stakeholder-aligned, human-centered benchmark framework that emphasizes explainability, value alignment, co-adaptive refinement, and iterative assessment\cite{han2023design, friedman2002value, mittal2024comprehensive}. We consider stakeholder aligned human-centered design and evaluation as an approach that centers educational design around humans' needs, engagement, and cognitive growth in this paper\cite{shaffer1999thick, shneiderman2022human,friedman2002value}.

\subsection{Foundational UX and Human-AI Evaluation Frameworks}

User experience (UX) evaluation plays a crucial role in assessing the effectiveness and acceptability of educational AI systems, particularly in human-centered contexts. Frameworks typically combine core evaluation criteria such as accuracy, clarity, feedback usefulness or engagement potential \cite{amershi2019guidelines, shute2013stealth, nielsen1995heuristics}. Human evaluation remains critical\cite{suchman2007human,thomas2024improving}. However, studies show that AI-generated feedback, although often perceived as an immediate assistant, also poses unique challenges related to explainability, bias, and ethical use \cite{liao2020questioning, binns2018s, bender2021dangers}; Trust, fairness, academic integrity, and need of AI literacy among educators and humans\cite{chaushi2023explainable, cotton2024chatting,long2020ai}. A notable limitation in existing UX evaluation research is its emphasis on domain specific systems, such as tutoring systems for STEAM like code.org\cite{grover2013computational}or writing tasks\cite{oneill2019stop}. Our work addresses this gap by proposing a domain independent UX evaluation framework that can generalize across creative, interdisciplinary learning environments such as Scratch, a block-based programming platform that supports storytelling, games\cite{resnick2015different}; Teachable Machine, train machine learning models through images, sounds\cite{zimmermann2019youth,touretzky2019special}, or Earsketch, expressive programming learning platform that support teaching both music and coding \cite{magerko2016earsketch}. These platforms support diverse, cross-disciplinary learning, but lack standardized frameworks to assess outcomes like adaptability, ethical awareness, human values, and stakeholder alignment. A flexible, domain independent benchmark is needed to capture the broader educational impact of these tools across varied contexts \cite{bommasani2021opportunities, xu2024theagentcompany}. In this paper, we define "Domain Independent Evaluation" as evaluating AI across multiple subject areas, requiring generalizable, content-neutral metrics\cite{reuel2024betterbench,bommasani2021opportunities}.

\subsection{Benchmarking and Evaluation of Intelligent Tutoring Systems}

For decades, traditional Intelligent Tutoring Systems (ITS) have aimed to deliver individualized instruction by modeling student knowledge and guiding problem-solving (inner loop) and instructional sequencing (outer loop) through predefined rule-based decision trees. These systems have shown positive learning outcomes through features such as immediate feedback and adaptivity \cite{vanlehn2006behavior, koedinger1997intelligent}. However, their reliance on rules limits responsiveness to dynamic learning scenarios and diverse student behaviors. In contrast, generative AI tutors are emerging to address these limitations by using adaptive techniques such as retrieval augmented generation for producing context-aware and coherent response\cite{liu2024can, lewis2020retrieval}, reinforcement learning to optimize teaching strategies based on students feedback \cite{chi2011evaluation}, deep knowledge tracing for modeling and predicting student understand over time\cite{piech2015deep}, and long-term retention and self-regulated learning over immediate correctness to support deeper learning \cite{shute2013stealth, roll2015understanding}.

Benchmarks are critical for evaluating educational AI systems, offering standardized tasks, datasets, and metrics to assess performance. In this context, we adopt the definition of benchmarking as “a combination of task, dataset, and metric” used to evaluate how AI systems support learning \cite{reuel2024betterbench, shute2013stealth}. But most existing benchmarks for large language models (LLMs) focus on general reasoning or factual recall \cite{bommasani2021opportunities}, with few targeting pedagogical efficacy in real-world learning contexts \cite{raji2021ai}. This gap raises concerns about the lack of stakeholder validation and limited alignment with teaching and learning needs and context \cite{reuel2024betterbench}. As Shute and Ventura emphasize, educational evaluation must move beyond correctness to include formative, contextual, and human-centered outcomes \cite{shute2013stealth, thomas2024improving}. Anderson et al. \cite{anderson1990cognitive} further demonstrated how AI tutors can be benchmarked to support procedural knowledge through structured feedback. Building on this, our work addresses the need for pedagogically grounded, stakeholder-aligned benchmarks that reflect how generative AI supports learning in authentic, situated contexts \cite{shaffer1999thick, friedman2002value}.

Our contribution to this paper is to address these research gaps through proposing the TEACH-AI (Trustworthy and Effective AI Classroom Heuristics) Benchmark Framework—a domain independent, value-aligned human-centered conceptual benchmark, along with a practical toolkit for evaluating generative AI tutors. Informed by a synthesis of over 126 publications, our framework serve as a start point to guide the design and evaluation of pedagogically meaningful AI-driven learning experiences.

\section{Methodology}

We conducted a scoping review following Arksey and O’Malley’s framework \cite{arksey2005scoping,levac2010scoping} to examine how AI agents are evaluated in educational environments. Guided by the question “How are AI agents evaluated in educational environments?”, we performed targeted searches across major venues (e.g., CHI, NeurIPS, IDC, AIED) and Google Scholar. In total, we reviewed 126 relevant sources, including \textbf{27 conference papers}, \textbf{78 journal articles}, and \textbf{21 books and gray literature}. These were categorized into three thematic phases: the \textbf{pre-LLM era} (pre-2017, focused on early ITS and HCI) with 37 papers, the \textbf{transformer era} (2017-2022, marked by the rise of XAI and AI literacy)\cite{vaswani2017attention, long2020ai} with 43 papers, and the \textbf{generative AI phase} (2023-present, emphasizing co-design and agent collaboration)\cite{denny2024generative, prasad2025exploring}, with 36 papers.

Through iterative coding and synthesis, we identified ten recurring components relevant to human-centered evaluation, including explainability, adaptivity, usability, ethical use, and accessibility. These insights informed a practical toolkit of reflective prompts \cite{wong2023seeing,friedman2002value} and a simplified scoring structure inspired by Meadows’ leverage points \cite{meadows1999leverage}. Regular weekly meetings with a senior faculty advisor facilitated thematic validation and iterative refinement of interpretations, ensuring conceptual rigor and alignment with human-centered AI evaluation principles in both the TEACH-AI benchmark and toolkit design.

\section{TEACH-AI Benchmark Framework and Early Design Implications}

In this section, we revisit our research question and present an initial benchmark framework along with a practical toolkit, drawing on existing literature, to address what evaluation components construct effective, value-aligned human–AI collaboration in educational domains. We define each component in detail and synthesize these findings into preliminary design implications to inform future benchmark development for generative AI tutoring agents. This benchmark framework adopts a value-sensitive human-centered perspective and structures the analysis to address the gap in existing evaluation approaches by strengthening the focus across cognitive and sociotechnical arguments and offers a foundation for iterative refinement through future research.

\subsection{TEACH-AI Evaluation Components}

To address the first part of the research question: What criteria define effective, value-aligned human–AI collaboration in education? We first define ten core components that form the basis of our evaluation framework (see Table~\ref{tab:evaluation_components}): explainability, helpfulness, adaptivity, consistency, creative exploration, system usability, ethical responsibility, accessibility, workflow, and refinement. We then provide a detailed table outlining sub-components with indicators or metrics, and relevant key references. 

\textbf{Explainability}: The agent’s ability to present its reasoning and decision making in clear, contextual meaningful, and human-understandable terms\cite{nauta2023anecdotal, arrieta2020explainable, guidotti2019stability}. 

\textbf{Helpfulness:} The extent to which the agent supports educational stakeholders such as teachers and humans' needs in achieving their goals through actionable, pedagogically appropriate assistance \cite{faruk2025introducing, puig2020watch, vygotsky1978mind}.

\textbf{Adaptivity:} The system’s responsiveness to human preferences, contexts, and needs through personalization and dynamic guidance. This includes flexible exploration to foster humans autonomy and confidence \cite{bransford2000people, gligorea2023adaptive, mulwa2011evaluation,nielsen1995heuristics}. 

\textbf{Consistency:} The stability and trustworthy of system outputs under similar conditions and alignment of behavior, languages, and situations across tasks \cite{nauta2023anecdotal, nielsen1995heuristics}.

\textbf{Learning Exploration:} The agent’s capacity to foster curiosity, support diverse solution paths, and encourage reflective, open-ended inquiry, long-term human autonomy \cite{shaffer1999thick, papert1993children, vygotsky1978mind, kapur2008productive,nielsen1995heuristics}.

\textbf{System Usability:} The effectiveness and ease of interaction that support efficient, intuitive, role-shifting, and error-resistant interactions between users and AI systems \cite{amershi2019guidelines, nielsen1995heuristics, turkle2011life}.

\textbf{Responsiblity and Ethics:} The system’s ability to act in alignment with human values, legal, ethical, and educational norms, and cultural sensitivities, even under adversarial conditions. It requires agents to avoid harm, ensure fairness, protect privacy and safeguard student data and voice \cite{wang2024ali, holmes2020artificial, bender2021dangers, lata2024beyond, bommasani2021opportunities}.

\textbf{Accessibility:} The extent to which the system is usable and equitable access by humans with diverse abilities, including those using assistive technologies \cite{world2012w3c, ruiz2024towards, prasad2025exploring, goldenthal2021not,ding2024redesigning}.

\textbf{Workflow Integration \& Stakeholder Coordination:} The agent’s ability to support multi-steps, human-AI collaboration between teachers, students, and other stakeholders, while maintaining adaptability in a dynamic learning context \cite{prasad2025exploring, amershi2019guidelines}.

\textbf{Refinement:} The system’s ability to support iterative improvement through a) the user correcting AI errors, b) user adjustment of vague or biased feedback, and c) ethical traceable revisions \cite{pan2024autonomous, wang2024ali,nielsen1995heuristics}.

Overall, TEACH-AI benchmark framework address three critical interconnected arguments in evaluation: (1) the agent’s capacity for explainability, adaptability, helpfulness, and consistency, including interpretable, context-aware justifications \cite{shaffer1999thick,bransford2000people}, dynamic adaptation to human needs \cite{puig2020watch,tack2022ai,bahel2024personalizing,brusilovsky2004layered}, and stable, reliable outputs scross similar conditions \cite{nielsen1995heuristics,vanden2012increasing};
(2) the extent to which the agent fosters creative exploration, emotional engagement, and deep thinking, by scaffolding open-ended problem-solving, supporting divergent approaches, encouraging productive struggle, and enabling transferable learning process cross domains\cite{amershi2019guidelines,koedinger1997intelligent,cherry2014quantifying, bransford2000people, nielsen1995heuristics}; and (3) the degree to which the agent operates responsibly, accessibly, and is open to refinement, including ethical behavior under adversarial conditions \cite{wang2024ali,bender2021dangers,bommasani2021opportunities,ding2025considering}, equitable access for diverse humans, and support for iterative improvement through feedback, error recovery, and coordination in multi-step, multi-stakeholder workflows \cite{nielsen1995heuristics,elnaggar2021quantification}.

\begin{table}[htbp]
\centering
\small
\caption{TEACH-AI Benchmark Framework: Evaluation Components for Generative AI Assistants}
\label{tab:evaluation_components}
\begin{adjustbox}{max width=\textwidth, center}
\begin{tabular}{>{\bfseries}p{2.5cm} p{4.0cm} p{8.0cm} p{3.0cm}}
\toprule
\textbf{Component} & \textbf{Subcomponents} & \textbf{Indicators / Metrics} & \textbf{Key References} \\
\midrule

1. Explainability & 
 Reasoning \newline  Clarity \newline  Traceability \newline  Fidelity \newline  Interpretability & 
XAI metrics, trust ratings, task agreement rate, XAI question bank & 
\cite{nauta2023anecdotal, rosenfeld2021better, silva2023explainable, ribera2019can, bommasani2021opportunities, bransford2000people} \par
\cite{abbas2022user, amershi2019guidelines, guidotti2018survey, liao2020questioning} \\ \addlinespace

2. Helpfulness & 
 Goal support \newline  human-aligned pedagogy & 
Task success rate, hint relevance (knowledge tracing), human modeling accuracy, user ratings (e.g., CAS-UX) & 
\cite{koedinger1997intelligent, puig2020watch, tack2022ai, bahel2024personalizing, faruk2025introducing} \\ \addlinespace

3. Adaptivity & 
 Personalization \newline  Context awareness \newline Controllability & 
Adaptation rate, System Usability Scale (SUS scores), User Experience Questionnaire (UEQ scores), NASA-TLX, controllability metrics, human modeling accuracy & 
\cite{brusilovsky2004layered, gligorea2023adaptive, perrig2024measurement, vanden2012increasing, bommasani2021opportunities} \par \cite{amershi2019guidelines, goldenthal2021not} \\ \addlinespace

4. Consistency & 
 Appropriate determinism  \newline Implementation invariance \newline Cross-evaluator reliability &
Output stability across runs, inter-coder agreement, threshold tuning precision, value map stability index& 
\cite{nauta2023anecdotal, carvalho2019machine, robnik2018perturbation, vanden2012increasing, nielsen1995heuristics} \\ \addlinespace

5. Learning Exploration & 
 Creativity \newline  Metacognition \newline  Transfer \newline  Affective \& social engagement & 
Creativity support index (CSI) scores, human modeling accuracy, transfer task performance, LX scale, Self-Determination Theory (SDT) indicators (autonomy, competence, relatedness) & 
\cite{shaffer1999thick, papert1993children, vygotsky1978mind, kapur2008productive, cherry2014quantifying} \par \cite{holstein2019co, safsouf2019design, ryan2000self, nielsen1995heuristics, prasad2025exploring} \\ \addlinespace

6. System Usability & 
Usability \newline Interface quality \newline Co-regulation support & 
CSI scores, usability heuristics checklist, interface clarity, feedback visibility, co-regulation cues rating(e.g., tutoring role-switching affordances), cognitive walkthrough analysis & 
\cite{cherry2014quantifying, turkle2011life, amershi2019guidelines, bevan1995human, nielsen1995heuristics, hadwin2011self,polson1992cognitive} \\ \addlinespace

7. Responsiblity \& Ethics & 
Fairness \newline Transparency \newline Privacy compliance & 
Fairness stress tests, adversarial prompt handling, stakeholder alignment, traceability, privacy compliance audit, group fairness metrics & 
\cite{wang2024ali, kilincc2024comprehensive, bommasani2021opportunities, holmes2020artificial, bender2021dangers, bellamy2019ai} \\ \addlinespace

8. Accessibility & 
Functional adaptation \newline Assistive technology integration \newline multi-modal UX compatibility & 
Text-to-audio adaptation rate, comprehension scores, error resolution rate, contextual navigability (e.g., keyboard and curriculum switching), XAI question bank, group fairness metrics & 
\cite{ruiz2024towards, morris2020ai, goldenthal2021not, mohamad2021ux, bellamy2019ai} \par \cite{goldsmith2007universal, world2012w3c, ding2025considering} \\ \addlinespace

9. Workflow \& Coordination & 
Multi-agent coordination \newline Multi-role flow control (teachers, students, stakeholders) & 
Workflow coherence score, task decomposition rate, planning cost analysis, human–agent alignment & 
\cite{xue2025comfybench, qiao2024benchmarking, xu2024theagentcompany, elnaggar2021quantification} \\ \addlinespace

10. Refinement & 
Iterative feedback \newline Error correction \newline Ethical traceability & 
Error detection rate for revisions, refinement trace logs (keystroke logs, edit history), cross-agent coherence, time-to-refine, value alignment over revisions, longitudinal user satisfaction & 
\cite{guo2024using, wang2024ali, pan2024autonomous, vasconcelos2022generation} \par  \cite{bryant2017automatic, hong2024my, jurenka2024towards} \\ \addlinespace

\bottomrule
\end{tabular}
\end{adjustbox}
\end{table}

\subsection{TEACH-AI Benchmark: Early Implications}

To illustrate how TEACH-AI can inform early-stage evaluation, we outline how the TEACH-AI framework could be applied to evaluate domain-independent generative AI assistants in educational settings \cite{dunne2024speculative}. The framework’s ten components can be selectively applied depending on research goals, stakeholder roles, and contextual factors. For instance, studies involving a single agent may emphasize components such as helpfulness or explainability, whereas multi-agent settings may prioritize coordination or workflow support. Similarly, accessibility considerations should be adapted based on the characteristics and needs of the target user population.

More broadly, TEACH-AI encourages researchers and designers to reflect on how generative AI systems support education, creativity, values, and human agency. By applying the framework iteratively, practitioners can identify where the system meets expectations and where further refinement is needed, guiding more thoughtful and contextually grounded algorithmic design decisions.

\subsection{Practical Toolkit} 
\subsubsection{TEACH-AI Toolkit Development: Checklist Example}

We also introduce a preliminary toolkit intended to help practitioners apply TEACH-AI in practice. The toolkit offers a set of reflective questions aligned with each framework component, supporting structured evaluation across different educational and design contexts. Rather than serving as a prescriptive checklist, these prompts help users identify strengths, gaps, and opportunities for improvement in an AI system’s behavior and alignment with human-centered values. The goal of this tool is to guide consistent reflection and comparison across contexts, whether in classroom use, design, or model development reviews, or early research prototyping. Future iterations will refine these prompts and explore ways to support broader, scalable evaluation workflows. This checklist can be used by educators, researchers, and designers to assess human-centered AI alignment.

\begin{table}[h]
\centering
\caption{TEACH-AI Toolkit Checklist Example (Index + Tech-Eval depth forthcoming).}
\small
\renewcommand{\arraystretch}{1.05}
\begin{tabular}{p{0.3\linewidth} p{0.62\linewidth}}
\hline
\textbf{Framework} & \textbf{Checklist Questions} \\
\hline

%Explainability & 
%a. Does the AI explain its decisions? b. Are sources provided? \\

%Learning Exploration & 
%a. Does it foster creativity, learning transfer, and critical thinking? b. Is feedback aligned with context(e.g., curriculum goals)? \\

%Responsiblity \& Ethics & 
%a. Does it support diverse users? b. Has it been tested for bias or adversarial prompts? c. Is data use (collection, storage, sharing) clear? d. Was a safety or privacy audit conducted?\\

\textbf{1. Explainability} &
a. Does the AI explain its decisions clearly? \newline
b. How would you verify the explanation? \\
\hline

\textbf{2. Helpfulness} &
a. Does the AI support the task goal you set?  \\
\hline

\textbf{3. Adaptivity} &
a. How quickly does the AI provide adaptive feedback (e.g., latency)?  \\
\hline

\textbf{4. Consistency} &
a. Does the system behave consistently across different conditions, prompts, or contexts?  \\
\hline

\textbf{5. Learning Exploration} &
a. Does the AI foster creativity, learning transfer, and critical thinking? \newline
b. Does the AI foster emotional connection, motivation, and social reasoning? \\
\hline

\textbf{6. System Usability} &
a. Was the AI easy to use and navigate?  \\
\hline

\textbf{7. Responsibility \& Ethics} &
a. Does the agent support diverse users equitably? \newline
b. Has a safety or privacy audit been conducted? \\
\hline

\textbf{8. Accessibility} &
a. Does the system adapt effectively to my accessibility needs?  \\
\hline

\textbf{9. Workflow \& Coordination} &
a. Were roles and responsibilities clear throughout the collaboration process? \\
\hline

\textbf{10. Refinement} &
a. Does the feedback clearly identify areas that need correction? \\
\hline
\end{tabular}

%\hline
%\end{tabular}
%\caption{Checklist Example: Applying the TEACH-AI Framework (expanded version forthcoming).}
\end{table}

%\subsubsection{Toolkit in Hypothetical Use Case}

 The checklist is intended to support reflective practice rather than function as a prescriptive to-do list. It translates abstract values (e.g., explainability) into actionable criteria that can be applied across technical design, policymaking, training, and research contexts \cite{friedman2019value}. In classroom settings, including those using tools like ChatGPT, the checklist guide scalable evaluation by allowing raters to assess each criterion using either a simple Yes/No option or a progressive scale \cite{meadows1999leverage}. This approach provides a clear foundation for assessing an AI system’s alignment with human values, contextual demands, and broad AI development principles such as transparency and safety across diverse educational environments. The resulting \textit{TEACH-AI index} can support both reflective classroom practice and quantitative research analysis, supporting consistent comparisons and guiding ongoing model evaluation efforts in the educational domain.

\section{Conclusion and future work}

In summary, we introduce TEACH-AI, a ten-component, human-centered benchmark framework and toolkit for evaluating generative AI systems in education. While the current version is primarily conceptual, it highlights the need for evaluation approaches that align with emerging educational needs, ethical design principles, and human values. Importantly, TEACH-AI bridges human-generated feedback and LLM-generated feedback by providing a unified structure that supports both human evaluators and LLM-as-judge methods.

Moving forward, our work will involve co-design with diverse stakeholders and iterative refinement of the framework across different educational contexts. We also plan to explore technical development, such as integrating TEACH-AI into a scalable digital prototype for large-scale benchmarking. This direction aligns with broader trends in human-centered AI evaluation, for example, the use of LLM-as-judge methods for automated assessment, and research on Reinforcement Learning from AI Feedback (RLAIF), which highlights the growing emphasis on reliable feedback signals in AI behavior. Our long-term goal is to support the development of accessible, responsible, and pedagogically aligned AI evaluation ecosystems that drive meaningful impact in real educational settings.

\section{References}

\bibliographystyle{unsrt}
\bibliography{refs}

\begin{thebibliography}{100}

\bibitem{shetye2024evaluation}
Shamini Shetye.
\newblock An evaluation of khanmigo, a generative ai tool, as a computer-assisted language learning app.
\newblock {\em Studies in Applied Linguistics and TESOL}, 24(1), 2024.

\bibitem{thomas2024using}
Danielle~R Thomas, Erin Gatz, Shivang Gupta, Jionghao Lin, Cindy Tipper, and Kenneth~R Koedinger.
\newblock Using generative ai to provide feedback to adult tutors in training and assess real-life performance.
\newblock In {\em The Learning Ideas Conference}, pages 204--214. Springer, 2024.

\bibitem{sperrle2021survey}
Fabian Sperrle, Mennatallah El-Assady, Grace Guo, Rita Borgo, D~Horng Chau, Alex Endert, and Daniel Keim.
\newblock A survey of human-centered evaluations in human-centered machine learning.
\newblock In {\em Computer Graphics Forum}, volume~40, pages 543--568. Wiley Online Library, 2021.

\bibitem{vanlehn2006behavior}
Kurt VanLehn.
\newblock The behavior of tutoring systems.
\newblock {\em International journal of artificial intelligence in education}, 16(3):227--265, 2006.

\bibitem{resnick2024generative}
Mitchel Resnick.
\newblock Generative ai and creative learning: Concerns, opportunities, and choices.
\newblock 2024.

\bibitem{shaffer1999thick}
David~Williamson Shaffer and Mitchel Resnick.
\newblock " thick" authenticity: New media and authentic learning.
\newblock {\em Journal of interactive learning research}, 10(2):195--216, 1999.

\bibitem{friedman2002value}
Batya Friedman, Peter Kahn, and Alan Borning.
\newblock Value sensitive design: Theory and methods.
\newblock {\em University of Washington technical report}, 2(8):1--8, 2002.

\bibitem{mcklin2018authenticity}
Tom McKlin, Brian Magerko, Taneisha Lee, Dana Wanzer, Doug Edwards, and Jason Freeman.
\newblock Authenticity and personal creativity: How earsketch affects student persistence.
\newblock In {\em Proceedings of the 49th ACM Technical Symposium on Computer Science Education}, pages 987--992, 2018.

\bibitem{ali2021social}
Safinah Ali, Hae~Won Park, and Cynthia Breazeal.
\newblock A social robot’s influence on children’s figural creativity during gameplay.
\newblock {\em International Journal of Child-Computer Interaction}, 28:100234, 2021.

\bibitem{vincent2020trustworthy}
St{\'e}phan Vincent-Lancrin and Reyer Van~der Vlies.
\newblock Trustworthy artificial intelligence (ai) in education: Promises and challenges.
\newblock {\em OECD education working papers}, (218):0\_1--17, 2020.

\bibitem{eshet2004digital}
Yoram Eshet.
\newblock Digital literacy: A conceptual framework for survival skills in the digital era.
\newblock {\em Journal of educational multimedia and hypermedia}, 13(1):93--106, 2004.

\bibitem{bundy2017preparing}
Alan Bundy.
\newblock Preparing for the future of artificial intelligence, 2017.

\bibitem{baer1998case}
John Baer.
\newblock The case for domain specificity of creativity.
\newblock {\em Creativity research journal}, 11(2):173--177, 1998.

\bibitem{amabile2018creativity}
Teresa~M Amabile.
\newblock {\em Creativity in context: Update to the social psychology of creativity}.
\newblock Routledge, 2018.

\bibitem{grover2013computational}
Shuchi Grover and Roy Pea.
\newblock Computational thinking in k--12: A review of the state of the field.
\newblock {\em Educational researcher}, 42(1):38--43, 2013.

\bibitem{vygotsky1978mind}
Lev~Semenovich Vygotsky and Michael Cole.
\newblock {\em Mind in society: Development of higher psychological processes}.
\newblock Harvard university press, 1978.

\bibitem{papert1991situating}
Seymour Papert and Idit Harel.
\newblock Situating constructionism.
\newblock {\em constructionism}, 36(2):1--11, 1991.

\bibitem{alasadi2023generative}
Eman~A Alasadi and Carlos~R Baiz.
\newblock Generative ai in education and research: Opportunities, concerns, and solutions.
\newblock {\em Journal of Chemical Education}, 100(8):2965--2971, 2023.

\bibitem{ash2020reimagining}
Katherine Ash and Madelyn Rahn.
\newblock Reimagining workforce policy in the age of disruption: A state guide for preparing the future workforce now.
\newblock {\em National Governors Association}, 2020.

\bibitem{friedman2019value}
Batya Friedman and David~G Hendry.
\newblock {\em Value sensitive design: Shaping technology with moral imagination}.
\newblock Mit Press, 2019.

\bibitem{han2023design}
Ariel Han and Zhenyao Cai.
\newblock Design implications of generative ai systems for visual storytelling for young learners.
\newblock In {\em Proceedings of the 22nd annual ACM interaction design and children conference}, pages 470--474, 2023.

\bibitem{mittal2024comprehensive}
Uday Mittal, Siva Sai, Vinay Chamola, et~al.
\newblock A comprehensive review on generative ai for education.
\newblock {\em IEEE Access}, 2024.

\bibitem{shneiderman2022human}
Ben Shneiderman.
\newblock {\em Human-centered AI}.
\newblock Oxford University Press, 2022.

\bibitem{amershi2019guidelines}
Saleema Amershi, Dan Weld, Mihaela Vorvoreanu, Adam Fourney, Besmira Nushi, Penny Collisson, Jina Suh, Shamsi Iqbal, Paul~N Bennett, Kori Inkpen, et~al.
\newblock Guidelines for human-ai interaction.
\newblock In {\em Proceedings of the 2019 chi conference on human factors in computing systems}, pages 1--13, 2019.

\bibitem{shute2013stealth}
Valerie Shute and Matthew Ventura.
\newblock {\em Stealth assessment: Measuring and supporting learning in video games}.
\newblock The mit press, 2013.

\bibitem{nielsen1995heuristics}
Jakob Nielsen.
\newblock Ten usability heuristics for user interface design.
\newblock Online; accessed July X, 2025, 1995.

\bibitem{suchman2007human}
Lucille~Alice Suchman.
\newblock {\em Human-machine reconfigurations: Plans and situated actions}.
\newblock Cambridge university press, 2007.

\bibitem{thomas2024improving}
Danielle~R Thomas, Jionghao Lin, Erin Gatz, Ashish Gurung, Shivang Gupta, Kole Norberg, Stephen~E Fancsali, Vincent Aleven, Lee Branstetter, Emma Brunskill, et~al.
\newblock Improving student learning with hybrid human-ai tutoring: A three-study quasi-experimental investigation.
\newblock In {\em Proceedings of the 14th Learning Analytics and Knowledge Conference}, pages 404--415, 2024.

\bibitem{liao2020questioning}
Q~Vera Liao, Daniel Gruen, and Sarah Miller.
\newblock Questioning the ai: informing design practices for explainable ai user experiences.
\newblock In {\em Proceedings of the 2020 CHI conference on human factors in computing systems}, pages 1--15, 2020.

\bibitem{binns2018s}
Reuben Binns, Max Van~Kleek, Michael Veale, Ulrik Lyngs, Jun Zhao, and Nigel Shadbolt.
\newblock 'it's reducing a human being to a percentage' perceptions of justice in algorithmic decisions.
\newblock In {\em Proceedings of the 2018 Chi conference on human factors in computing systems}, pages 1--14, 2018.

\bibitem{bender2021dangers}
Emily~M Bender, Timnit Gebru, Angelina McMillan-Major, and Shmargaret Shmitchell.
\newblock On the dangers of stochastic parrots: Can language models be too big?
\newblock In {\em Proceedings of the 2021 ACM conference on fairness, accountability, and transparency}, pages 610--623, 2021.

\bibitem{chaushi2023explainable}
Blerta~Abazi Chaushi, Besnik Selimi, Agron Chaushi, and Marika Apostolova.
\newblock Explainable artificial intelligence in education: A comprehensive review.
\newblock In {\em World Conference on Explainable Artificial Intelligence}, pages 48--71. Springer, 2023.

\bibitem{cotton2024chatting}
Debby~RE Cotton, Peter~A Cotton, and J~Reuben Shipway.
\newblock Chatting and cheating: Ensuring academic integrity in the era of chatgpt.
\newblock {\em Innovations in education and teaching international}, 61(2):228--239, 2024.

\bibitem{long2020ai}
Duri Long and Brian Magerko.
\newblock What is ai literacy? competencies and design considerations.
\newblock In {\em Proceedings of the 2020 CHI conference on human factors in computing systems}, pages 1--16, 2020.

\bibitem{oneill2019stop}
Ruth ONeill and Alex Russell.
\newblock Stop! grammar time: University students’ perceptions of the automated feedback program grammarly.
\newblock {\em Australasian Journal of Educational Technology}, 35(1), 2019.

\bibitem{resnick2015different}
Mitchel Resnick and David Siegel.
\newblock A different approach to coding.
\newblock {\em International Journal of People-Oriented Programming}, 4(1):1--4, 2015.

\bibitem{zimmermann2019youth}
Abigail Zimmermann-Niefield, Makenna Turner, Bridget Murphy, Shaun~K Kane, and R~Benjamin Shapiro.
\newblock Youth learning machine learning through building models of athletic moves.
\newblock In {\em Proceedings of the 18th ACM international conference on interaction design and children}, pages 121--132, 2019.

\bibitem{touretzky2019special}
David Touretzky, Fred Martin, Deborah Seehorn, Cynthia Breazeal, and Tess Posner.
\newblock Special session: Ai for k-12 guidelines initiative.
\newblock In {\em Proceedings of the 50th ACM technical symposium on computer science education}, pages 492--493, 2019.

\bibitem{magerko2016earsketch}
Brian Magerko, Jason Freeman, Tom Mcklin, Mike Reilly, Elise Livingston, Scott Mccoid, and Andrea Crews-Brown.
\newblock Earsketch: A steam-based approach for underrepresented populations in high school computer science education.
\newblock {\em ACM Transactions on Computing Education (TOCE)}, 16(4):1--25, 2016.

\bibitem{bommasani2021opportunities}
Rishi Bommasani, Drew~A Hudson, Ehsan Adeli, Russ Altman, Simran Arora, Sydney von Arx, Michael~S Bernstein, Jeannette Bohg, Antoine Bosselut, Emma Brunskill, et~al.
\newblock On the opportunities and risks of foundation models.
\newblock {\em arXiv preprint arXiv:2108.07258}, 2021.

\bibitem{xu2024theagentcompany}
Frank~F Xu, Yufan Song, Boxuan Li, Yuxuan Tang, Kritanjali Jain, Mengxue Bao, Zora~Z Wang, Xuhui Zhou, Zhitong Guo, Murong Cao, et~al.
\newblock Theagentcompany: benchmarking llm agents on consequential real world tasks.
\newblock {\em arXiv preprint arXiv:2412.14161}, 2024.

\bibitem{reuel2024betterbench}
Anka Reuel-Lamparth, Amelia Hardy, Chandler Smith, Max Lamparth, Malcolm Hardy, and Mykel~J Kochenderfer.
\newblock Betterbench: Assessing ai benchmarks, uncovering issues, and establishing best practices.
\newblock {\em Advances in Neural Information Processing Systems}, 37:21763--21813, 2024.

\bibitem{koedinger1997intelligent}
Kenneth~R Koedinger, John~R Anderson, William~H Hadley, and Mary~A Mark.
\newblock Intelligent tutoring goes to school in the big city.
\newblock {\em International Journal of Artificial Intelligence in Education}, 8:30--43, 1997.

\bibitem{liu2024can}
Suqing Liu, Zezhu Yu, Feiran Huang, Yousef Bulbulia, Andreas Bergen, and Michael Liut.
\newblock Can small language models with retrieval-augmented generation replace large language models when learning computer science?
\newblock In {\em Proceedings of the 2024 on Innovation and Technology in Computer Science Education V. 1}, pages 388--393. 2024.

\bibitem{lewis2020retrieval}
Patrick Lewis, Ethan Perez, Aleksandra Piktus, Fabio Petroni, Vladimir Karpukhin, Naman Goyal, Heinrich K{\"u}ttler, Mike Lewis, Wen-tau Yih, Tim Rockt{\"a}schel, et~al.
\newblock Retrieval-augmented generation for knowledge-intensive nlp tasks.
\newblock {\em Advances in neural information processing systems}, 33:9459--9474, 2020.

\bibitem{chi2011evaluation}
Min Chi, Kurt VanLehn, Diane Litman, and Pamela Jordan.
\newblock An evaluation of pedagogical tutorial tactics for a natural language tutoring system: A reinforcement learning approach.
\newblock {\em International Journal of Artificial Intelligence in Education}, 21(1-2):83--113, 2011.

\bibitem{piech2015deep}
Chris Piech, Jonathan Bassen, Jonathan Huang, Surya Ganguli, Mehran Sahami, Leonidas~J Guibas, and Jascha Sohl-Dickstein.
\newblock Deep knowledge tracing.
\newblock {\em Advances in neural information processing systems}, 28, 2015.

\bibitem{roll2015understanding}
Ido Roll and Philip~H Winne.
\newblock Understanding, evaluating, and supporting self-regulated learning using learning analytics.
\newblock {\em Journal of Learning Analytics}, 2(1):7--12, 2015.

\bibitem{raji2021ai}
Inioluwa~Deborah Raji, Emily~M Bender, Amandalynne Paullada, Emily Denton, and Alex Hanna.
\newblock Ai and the everything in the whole wide world benchmark.
\newblock {\em arXiv preprint arXiv:2111.15366}, 2021.

\bibitem{anderson1990cognitive}
John~R Anderson, C~Franklin Boyle, Albert~T Corbett, and Matthew~W Lewis.
\newblock Cognitive modeling and intelligent tutoring.
\newblock {\em Artificial intelligence}, 42(1):7--49, 1990.

\bibitem{arksey2005scoping}
Hilary Arksey and Lisa O'malley.
\newblock Scoping studies: towards a methodological framework.
\newblock {\em International journal of social research methodology}, 8(1):19--32, 2005.

\bibitem{levac2010scoping}
Danielle Levac, Heather Colquhoun, and Kelly~K O'brien.
\newblock Scoping studies: advancing the methodology.
\newblock {\em Implementation science}, 5:1--9, 2010.

\bibitem{vaswani2017attention}
Ashish Vaswani, Noam Shazeer, Niki Parmar, Jakob Uszkoreit, Llion Jones, Aidan~N Gomez, {\L}ukasz Kaiser, and Illia Polosukhin.
\newblock Attention is all you need.
\newblock {\em Advances in neural information processing systems}, 30, 2017.

\bibitem{denny2024generative}
Paul Denny, Sumit Gulwani, Neil~T Heffernan, Tanja K{\"a}ser, Steven Moore, Anna~N Rafferty, and Adish Singla.
\newblock Generative ai for education (gaied): Advances, opportunities, and challenges.
\newblock {\em arXiv preprint arXiv:2402.01580}, 2024.

\bibitem{prasad2025exploring}
Prajish Prasad, Rishabh Balse, and Dhwani Balchandani.
\newblock Exploring multimodal generative ai for education through co-design workshops with students.
\newblock In {\em Proceedings of the 2025 CHI Conference on Human Factors in Computing Systems}, pages 1--17, 2025.

\bibitem{wong2023seeing}
Richmond~Y Wong, Michael~A Madaio, and Nick Merrill.
\newblock Seeing like a toolkit: How toolkits envision the work of ai ethics.
\newblock {\em Proceedings of the ACM on Human-Computer Interaction}, 7(CSCW1):1--27, 2023.

\bibitem{meadows1999leverage}
Donella Meadows.
\newblock Leverage points.
\newblock {\em Places to Intervene in a System}, 19:28, 1999.

\bibitem{nauta2023anecdotal}
Meike Nauta, Jan Trienes, Shreyasi Pathak, Elisa Nguyen, Michelle Peters, Yasmin Schmitt, J{\"o}rg Schl{\"o}tterer, Maurice Van~Keulen, and Christin Seifert.
\newblock From anecdotal evidence to quantitative evaluation methods: A systematic review on evaluating explainable ai.
\newblock {\em ACM Computing Surveys}, 55(13s):1--42, 2023.

\bibitem{arrieta2020explainable}
Alejandro~Barredo Arrieta, Natalia D{\'\i}az-Rodr{\'\i}guez, Javier Del~Ser, Adrien Bennetot, Siham Tabik, Alberto Barbado, Salvador Garc{\'\i}a, Sergio Gil-L{\'o}pez, Daniel Molina, Richard Benjamins, et~al.
\newblock Explainable artificial intelligence (xai): Concepts, taxonomies, opportunities and challenges toward responsible ai.
\newblock {\em Information fusion}, 58:82--115, 2020.

\bibitem{guidotti2019stability}
Riccardo Guidotti and Salvatore Ruggieri.
\newblock On the stability of interpretable models.
\newblock In {\em 2019 international joint conference on neural networks (IJCNN)}, pages 1--8. IEEE, 2019.

\bibitem{faruk2025introducing}
Lawal Ibrahim~Dutsinma Faruk, Debajyoti Pal, Suree Funilkul, Thinagaran Perumal, and Pornchai Mongkolnam.
\newblock Introducing casux: A standardized scale for measuring the user experience of artificial intelligence based conversational agents.
\newblock {\em International Journal of Human--Computer Interaction}, 41(9):5274--5298, 2025.

\bibitem{puig2020watch}
Xavier Puig, Tianmin Shu, Shuang Li, Zilin Wang, Yuan-Hong Liao, Joshua~B Tenenbaum, Sanja Fidler, and Antonio Torralba.
\newblock Watch-and-help: A challenge for social perception and human-ai collaboration.
\newblock {\em arXiv preprint arXiv:2010.09890}, 2020.

\bibitem{bransford2000people}
John~D Bransford, Ann~L Brown, Rodney~R Cocking, et~al.
\newblock {\em How people learn}, volume~11.
\newblock Washington, DC: National academy press, 2000.

\bibitem{gligorea2023adaptive}
Ilie Gligorea, Marius Cioca, Romana Oancea, Andra-Teodora Gorski, Hortensia Gorski, and Paul Tudorache.
\newblock Adaptive learning using artificial intelligence in e-learning: A literature review.
\newblock {\em Education Sciences}, 13(12):1216, 2023.

\bibitem{mulwa2011evaluation}
Catherine Mulwa, Seamus Lawless, Mary Sharp, and Vincent Wade.
\newblock The evaluation of adaptive and personalised information retrieval systems: a review.
\newblock {\em International Journal of Knowledge and Web Intelligence}, 2(2-3):138--156, 2011.

\bibitem{papert1993children}
Seymour Papert.
\newblock {\em The children's machine: Rethinking school in the age of the computer}.
\newblock Basic Books, Inc., 1993.

\bibitem{kapur2008productive}
Manu Kapur.
\newblock Productive failure.
\newblock {\em Cognition and instruction}, 26(3):379--424, 2008.

\bibitem{turkle2011life}
Sherry Turkle.
\newblock {\em Life on the Screen}.
\newblock Simon and Schuster, 2011.

\bibitem{wang2024ali}
Han Wang, An~Zhang, Nguyen Duy~Tai, Jun Sun, Tat-Seng Chua, et~al.
\newblock Ali-agent: Assessing llms' alignment with human values via agent-based evaluation.
\newblock {\em Advances in Neural Information Processing Systems}, 37:99040--99088, 2024.

\bibitem{holmes2020artificial}
Wayne Holmes.
\newblock Artificial intelligence in education.
\newblock In {\em Encyclopedia of education and information technologies}, pages 88--103. Springer, 2020.

\bibitem{lata2024beyond}
Prem Lata.
\newblock Beyond algorithms: Humanizing artificial intelligence for personalized and adaptive learning.
\newblock {\em International Journal of Innovative Research in Engineering and Management}, 11(5):10--55524, 2024.

\bibitem{world2012w3c}
World Wide~Web Consortium et~al.
\newblock W3c web content accessibility guidelines (wcag 2.0).
\newblock {\em Internet]. World Wide Web Consortium. Accessed}, 22, 2012.

\bibitem{ruiz2024towards}
Diana Ruiz and Tom Duenas.
\newblock Towards inclusive ai: Developing a w3c-inspired accessibility benchmark for large language models.
\newblock {\em Research Gate}, 2024.

\bibitem{goldenthal2021not}
Emma Goldenthal, Jennifer Park, Sunny~X Liu, Hannah Mieczkowski, and Jeffrey~T Hancock.
\newblock Not all ai are equal: Exploring the accessibility of ai-mediated communication technology.
\newblock {\em Computers in Human Behavior}, 125:106975, 2021.

\bibitem{ding2024redesigning}
Shi Ding, Jason~Brent Smith, Stephen Garrett, and Brian Magerko.
\newblock Redesigning earsketch for inclusive cs education: A participatory design approach.
\newblock In {\em Proceedings of the 23rd Annual ACM Interaction Design and Children Conference}, pages 720--724, 2024.

\bibitem{pan2024autonomous}
Jiayi Pan, Yichi Zhang, Nicholas Tomlin, Yifei Zhou, Sergey Levine, and Alane Suhr.
\newblock Autonomous evaluation and refinement of digital agents.
\newblock {\em arXiv preprint arXiv:2404.06474}, 2024.

\bibitem{tack2022ai}
Ana{\"\i}s Tack and Chris Piech.
\newblock The ai teacher test: Measuring the pedagogical ability of blender and gpt-3 in educational dialogues.
\newblock {\em arXiv preprint arXiv:2205.07540}, 2022.

\bibitem{bahel2024personalizing}
Vedant Bahel, Harshinee Sriram, and Cristina Conati.
\newblock Personalizing explanations of ai-driven hints to users' cognitive abilities: an empirical evaluation.
\newblock {\em arXiv preprint arXiv:2403.04035}, 2024.

\bibitem{brusilovsky2004layered}
Peter Brusilovsky, Charalampos Karagiannidis, and Demetrios Sampson.
\newblock Layered evaluation of adaptive learning systems.
\newblock {\em International Journal of Continuing Engineering Education and Life Long Learning}, 14(4-5):402--421, 2004.

\bibitem{vanden2012increasing}
Vero Vanden~Abeele, Erik Hauters, and Bieke Zaman.
\newblock Increasing the reliability and validity of quantitative laddering data with ladderux.
\newblock In {\em CHI'12 Extended Abstracts on Human Factors in Computing Systems}, pages 2057--2062. 2012.

\bibitem{cherry2014quantifying}
Erin Cherry and Celine Latulipe.
\newblock Quantifying the creativity support of digital tools through the creativity support index.
\newblock {\em ACM Transactions on Computer-Human Interaction (TOCHI)}, 21(4):1--25, 2014.

\bibitem{ding2025considering}
Shi Ding, Jason~Brent Smith, and Brian Magerko.
\newblock Considering large language model integration in expressive computer science learning environments for blind and visually impaired learners through co-design.
\newblock In {\em International Conference on Artificial Intelligence in Education}, pages 472--480. Springer, 2025.

\bibitem{elnaggar2021quantification}
Omar Elnaggar and Roselina Arelhi.
\newblock Quantification of knowledge exchange within classrooms: an ai-based approach.
\newblock In {\em The European Conference on Education}, pages 1--11, 2021.

\bibitem{rosenfeld2021better}
Avi Rosenfeld.
\newblock Better metrics for evaluating explainable artificial intelligence.
\newblock In {\em Proceedings of the 20th international conference on autonomous agents and multiagent systems}, pages 45--50, 2021.

\bibitem{silva2023explainable}
Andrew Silva, Mariah Schrum, Erin Hedlund-Botti, Nakul Gopalan, and Matthew Gombolay.
\newblock Explainable artificial intelligence: Evaluating the objective and subjective impacts of xai on human-agent interaction.
\newblock {\em International Journal of Human--Computer Interaction}, 39(7):1390--1404, 2023.

\bibitem{ribera2019can}
Mireia Ribera and Agata Lapedriza.
\newblock Can we do better explanations? a proposal of user-centered explainable ai.
\newblock CEUR Workshop Proceedings, 2019.

\bibitem{abbas2022user}
Abdallah~MH Abbas, Khairil~Imran Ghauth, and Choo-Yee Ting.
\newblock User experience design using machine learning: a systematic review.
\newblock {\em IEEE Access}, 10:51501--51514, 2022.

\bibitem{guidotti2018survey}
Riccardo Guidotti, Anna Monreale, Salvatore Ruggieri, Franco Turini, Fosca Giannotti, and Dino Pedreschi.
\newblock A survey of methods for explaining black box models.
\newblock {\em ACM computing surveys (CSUR)}, 51(5):1--42, 2018.

\bibitem{perrig2024measurement}
Sebastian~AC Perrig, Lena~Fanya Aeschbach, Nicolas Scharowski, Nick von Felten, Klaus Opwis, and Florian Br{\"u}hlmann.
\newblock Measurement practices in user experience (ux) research: A systematic quantitative literature review.
\newblock {\em Frontiers in Computer Science}, 6:1368860, 2024.

\bibitem{carvalho2019machine}
Diogo~V Carvalho, Eduardo~M Pereira, and Jaime~S Cardoso.
\newblock Machine learning interpretability: A survey on methods and metrics.
\newblock {\em Electronics}, 8(8):832, 2019.

\bibitem{robnik2018perturbation}
Marko Robnik-{\v{S}}ikonja and Marko Bohanec.
\newblock Perturbation-based explanations of prediction models.
\newblock In {\em Human and Machine Learning: Visible, Explainable, Trustworthy and Transparent}, pages 159--175. Springer, 2018.

\bibitem{holstein2019co}
Kenneth Holstein, Bruce~M McLaren, and Vincent Aleven.
\newblock Co-designing a real-time classroom orchestration tool to support teacher-ai complementarity.
\newblock {\em Grantee Submission}, 2019.

\bibitem{safsouf2019design}
Yassine Safsouf, Khalifa Mansouri, and Franck Poirier.
\newblock Design of a new scale to measure the learner experience in e-learning systems.
\newblock In {\em Proceedings of the International Conference on E-Learning}, pages 301--304. ERIC, 2019.

\bibitem{ryan2000self}
Richard~M Ryan and Edward~L Deci.
\newblock Self-determination theory and the facilitation of intrinsic motivation, social development, and well-being.
\newblock {\em American psychologist}, 55(1):68, 2000.

\bibitem{bevan1995human}
Nigel Bevan.
\newblock Human-computer interaction standards.
\newblock In {\em Advances in human factors/ergonomics}, volume~20, pages 885--890. Elsevier, 1995.

\bibitem{hadwin2011self}
Allyson~Fiona Hadwin, Sanna J{\"a}rvel{\"a}, and Mariel Miller.
\newblock Self-regulated, co-regulated, and socially shared regulation of learning.
\newblock {\em Handbook of self-regulation of learning and performance}, 30:65--84, 2011.

\bibitem{polson1992cognitive}
Peter~G Polson, Clayton Lewis, John Rieman, and Cathleen Wharton.
\newblock Cognitive walkthroughs: a method for theory-based evaluation of user interfaces.
\newblock {\em International Journal of man-machine studies}, 36(5):741--773, 1992.

\bibitem{kilincc2024comprehensive}
Sel{\c{c}}uk K{\i}l{\i}n{\c{c}}.
\newblock Comprehensive ai assessment framework: Enhancing educational evaluation with ethical ai integration.
\newblock {\em Journal of Educational Technology and Online Learning}, 7(4-ICETOL 2024 Special Issue):521--540, 2024.

\bibitem{bellamy2019ai}
Rachel~KE Bellamy, Kuntal Dey, Michael Hind, Samuel~C Hoffman, Stephanie Houde, Kalapriya Kannan, Pranay Lohia, Jacquelyn Martino, Sameep Mehta, Aleksandra Mojsilovi{\'c}, et~al.
\newblock Ai fairness 360: An extensible toolkit for detecting and mitigating algorithmic bias.
\newblock {\em IBM Journal of Research and Development}, 63(4/5):4--1, 2019.

\bibitem{morris2020ai}
Meredith~Ringel Morris.
\newblock Ai and accessibility.
\newblock {\em Communications of the ACM}, 63(6):35--37, 2020.

\bibitem{mohamad2021ux}
Normala Mohamad, Nor~Laily Hashim, Husna~Mad Baguri, Hazirah~Abdul Pisal, Cik~Fazilah Hibadullah, and Nur Hani~Zulkifli Abai.
\newblock Ux metrics of mobile learning for deaf children using fuzzy delphi method.
\newblock In {\em 2021 IEEE International Conference on Computing (ICOCO)}, pages 309--314. IEEE, 2021.

\bibitem{goldsmith2007universal}
Selwyn Goldsmith.
\newblock {\em Universal design}.
\newblock Routledge, 2007.

\bibitem{xue2025comfybench}
Xiangyuan Xue, Zeyu Lu, Di~Huang, Zidong Wang, Wanli Ouyang, and Lei Bai.
\newblock Comfybench: Benchmarking llm-based agents in comfyui for autonomously designing collaborative ai systems.
\newblock In {\em Proceedings of the Computer Vision and Pattern Recognition Conference}, pages 24614--24624, 2025.

\bibitem{qiao2024benchmarking}
Shuofei Qiao, Runnan Fang, Zhisong Qiu, Xiaobin Wang, Ningyu Zhang, Yong Jiang, Pengjun Xie, Fei Huang, and Huajun Chen.
\newblock Benchmarking agentic workflow generation.
\newblock {\em arXiv preprint arXiv:2410.07869}, 2024.

\bibitem{guo2024using}
Shuchen Guo, Ehsan Latif, Yifan Zhou, Xuan Huang, and Xiaoming Zhai.
\newblock Using generative ai and multi-agents to provide automatic feedback.
\newblock {\em arXiv preprint arXiv:2411.07407}, 2024.

\bibitem{vasconcelos2022generation}
Helena Vasconcelos, Gagan Bansal, Adam Fourney, Q~Vera Liao, and Jennifer~Wortman Vaughan.
\newblock Generation probabilities are not enough: Improving error highlighting for ai code suggestions.
\newblock In {\em HCAI Workshop at NeurIPS}, 2022.

\bibitem{bryant2017automatic}
CJ~Bryant, Mariano Felice, and Edward Briscoe.
\newblock Automatic annotation and evaluation of error types for grammatical error correction.
\newblock Association for Computational Linguistics, 2017.

\bibitem{hong2024my}
Shengxin Hong, Chang Cai, Sixuan Du, Haiyue Feng, Siyuan Liu, and Xiuyi Fan.
\newblock " my grade is wrong!": A contestable ai framework for interactive feedback in evaluating student essays.
\newblock {\em arXiv preprint arXiv:2409.07453}, 2024.

\bibitem{jurenka2024towards}
Irina Jurenka, Markus Kunesch, Kevin~R McKee, Daniel Gillick, Shaojian Zhu, Sara Wiltberger, Shubham~Milind Phal, Katherine Hermann, Daniel Kasenberg, Avishkar Bhoopchand, et~al.
\newblock Towards responsible development of generative ai for education: An evaluation-driven approach.
\newblock {\em arXiv preprint arXiv:2407.12687}, 2024.

\bibitem{dunne2024speculative}
Anthony Dunne and Fiona Raby.
\newblock {\em Speculative Everything, With a new preface by the authors: Design, Fiction, and Social Dreaming}.
\newblock MIT press, 2024.

\bibitem{rahimi2023validity}
Seyedahmad Rahimi, Jason~Brent Smith, Erin~JK Truesdell, Ashvala Vinay, Kristy~Elizabeth Boyer, Brian Magerko, Jason Freeman, and Tom Mcklin.
\newblock Validity and fairness of an automated assessment of creativity in computational music remixing.
\newblock In {\em AAGPW@ AIED}, pages 36--44, 2023.

\bibitem{brown2020language}
Tom Brown, Benjamin Mann, Nick Ryder, Melanie Subbiah, Jared~D Kaplan, Prafulla Dhariwal, Arvind Neelakantan, Pranav Shyam, Girish Sastry, Amanda Askell, et~al.
\newblock Language models are few-shot learners.
\newblock {\em Advances in neural information processing systems}, 33:1877--1901, 2020.

\bibitem{dignum2019responsible}
Virginia Dignum.
\newblock {\em Responsible artificial intelligence: how to develop and use AI in a responsible way}, volume 2156.
\newblock Springer, 2019.

\bibitem{schellaert2023your}
Wout Schellaert, Fernando Mart{\'\i}nez-Plumed, Karina Vold, John Burden, Pablo~AM Casares, Bao~Sheng Loe, Roi Reichart, Sean {\'O}~h{\'E}igeartaigh, Anna Korhonen, and Jos{\'e} Hern{\'a}ndez-Orallo.
\newblock Your prompt is my command: on assessing the human-centred generality of multimodal models.
\newblock {\em Journal of Artificial Intelligence Research}, 77:377--394, 2023.

\bibitem{lynch2016evaluation}
Tiina Lynch and Ioana Ghergulescu.
\newblock An evaluation framework for adaptive and intelligent tutoring systems.
\newblock In {\em E-learn: world conference on e-learning in corporate, government, healthcare, and higher education}, pages 1385--1390. Association for the Advancement of Computing in Education (AACE), 2016.

\bibitem{brooke1996sus}
John Brooke et~al.
\newblock Sus-a quick and dirty usability scale.
\newblock {\em Usability evaluation in industry}, 189(194):4--7, 1996.

\bibitem{laugwitz2008construction}
Bettina Laugwitz, Theo Held, and Martin Schrepp.
\newblock Construction and evaluation of a user experience questionnaire.
\newblock In {\em Symposium of the Austrian HCI and usability engineering group}, pages 63--76. Springer, 2008.

\bibitem{hart1988development}
Sandra~G Hart and Lowell~E Staveland.
\newblock Development of nasa-tlx (task load index): Results of empirical and theoretical research.
\newblock In {\em Advances in psychology}, volume~52, pages 139--183. Elsevier, 1988.

\bibitem{law2014attitudes}
Effie Lai-Chong Law, Paul Van~Schaik, and Virpi Roto.
\newblock Attitudes towards user experience (ux) measurement.
\newblock {\em International Journal of Human-Computer Studies}, 72(6):526--541, 2014.

\bibitem{bodria2023benchmarking}
Francesco Bodria, Fosca Giannotti, Riccardo Guidotti, Francesca Naretto, Dino Pedreschi, and Salvatore Rinzivillo.
\newblock Benchmarking and survey of explanation methods for black box models.
\newblock {\em Data Mining and Knowledge Discovery}, 37(5):1719--1778, 2023.

\bibitem{dweck2008brainology}
Carol~S Dweck.
\newblock Brainology: Transforming students’ motivation to learn.
\newblock {\em Independent school}, 67(2):110--119, 2008.

\bibitem{dweck2010mind}
Carol~S Dweck.
\newblock Mind-sets.
\newblock {\em Principal leadership}, 10(5):26--29, 2010.

\bibitem{koedinger2006cognitive}
Kenneth~R Koedinger, Albert Corbett, et~al.
\newblock {\em Cognitive tutors: Technology bringing learning sciences to the classroom}.
\newblock na, 2006.

\bibitem{collins2013cognitive}
Allan Collins.
\newblock Cognitive apprenticeship and instructional technology.
\newblock In {\em Educational values and cognitive instruction}, pages 121--138. Routledge, 2013.

\bibitem{mcarthur1990tutoring}
David McArthur, Cathleen Stasz, and Mary Zmuidzinas.
\newblock Tutoring techniques in algebra.
\newblock {\em Cognition and instruction}, 7(3):197--244, 1990.

\bibitem{vanlehn2011relative}
Kurt VanLehn.
\newblock The relative effectiveness of human tutoring, intelligent tutoring systems, and other tutoring systems.
\newblock {\em Educational psychologist}, 46(4):197--221, 2011.

\bibitem{luckin2016intelligence}
Rose Luckin and Wayne Holmes.
\newblock Intelligence unleashed: An argument for ai in education.
\newblock 2016.

\bibitem{kirschner2006unguided}
Paul Kirschner, John Sweller, and Richard~E Clark.
\newblock Why unguided learning does not work: An analysis of the failure of discovery learning, problem-based learning, experiential learning and inquiry-based learning.
\newblock {\em Educational psychologist}, 41(2):75--86, 2006.

\bibitem{piaget1954development}
Jean Piaget and Margaret~Trans Cook.
\newblock The development of object concept.
\newblock 1954.

\bibitem{nathan2007value}
Lisa~P Nathan, Predrag~V Klasnja, and Batya Friedman.
\newblock Value scenarios: a technique for envisioning systemic effects of new technologies.
\newblock In {\em CHI'07 extended abstracts on Human factors in computing systems}, pages 2585--2590, 2007.

\bibitem{schmidgall2024agentclinic}
Samuel Schmidgall, Rojin Ziaei, Carl Harris, Eduardo Reis, Jeffrey Jopling, and Michael Moor.
\newblock Agentclinic: a multimodal agent benchmark to evaluate ai in simulated clinical environments.
\newblock {\em arXiv preprint arXiv:2405.07960}, 2024.

\bibitem{hendrycks2020aligning}
Dan Hendrycks, Collin Burns, Steven Basart, Andrew Critch, Jerry Li, Dawn Song, and Jacob Steinhardt.
\newblock Aligning ai with shared human values.
\newblock {\em arXiv preprint arXiv:2008.02275}, 2020.

\bibitem{haidt2003moral}
Jonathan Haidt et~al.
\newblock The moral emotions.
\newblock {\em Handbook of affective sciences}, 11(2003):852--870, 2003.

\bibitem{goodfellow2014explaining}
Ian~J Goodfellow, Jonathon Shlens, and Christian Szegedy.
\newblock Explaining and harnessing adversarial examples.
\newblock {\em arXiv preprint arXiv:1412.6572}, 2014.

\bibitem{jian2000foundations}
Jiun-Yin Jian, Ann~M Bisantz, and Colin~G Drury.
\newblock Foundations for an empirically determined scale of trust in automated systems.
\newblock {\em International journal of cognitive ergonomics}, 4(1):53--71, 2000.

\bibitem{bartneck2009measurement}
Christoph Bartneck, Dana Kuli{\'c}, Elizabeth Croft, and Susana Zoghbi.
\newblock Measurement instruments for the anthropomorphism, animacy, likeability, perceived intelligence, and perceived safety of robots.
\newblock {\em International journal of social robotics}, 1(1):71--81, 2009.

\bibitem{barredo2019explainable}
Alejandro Barredo~Arrieta, F~Herrera, R~Chatilaf, R~Benjaminsh, D~Molina, S~Gil-Lopez, Salvador Garcia, A~Barbadoh, S~Tabikg, A~Bennetot, et~al.
\newblock Explainable artificial intelligence (xai): Concepts, taxonomies, opportunities and challenges toward responsible ai.
\newblock 2019.

\bibitem{kalbag2017accessibility}
Laura Kalbag and Heydon Pickering.
\newblock {\em Accessibility for everyone}.
\newblock A Book Apart New York, 2017.

\bibitem{holmes2018mismatch}
Kat Holmes.
\newblock {\em Mismatch: How inclusion shapes design}.
\newblock Mit Press, 2018.

\bibitem{nielsen1994enhancing}
Jakob Nielsen.
\newblock Enhancing the explanatory power of usability heuristics.
\newblock In {\em Proceedings of the SIGCHI conference on Human Factors in Computing Systems}, pages 152--158, 1994.

\bibitem{resnick2009scratch}
Mitchel Resnick, John Maloney, Andr{\'e}s Monroy-Hern{\'a}ndez, Natalie Rusk, Evelyn Eastmond, Karen Brennan, Amon Millner, Eric Rosenbaum, Jay Silver, Brian Silverman, et~al.
\newblock Scratch: programming for all.
\newblock {\em Communications of the ACM}, 52(11):60--67, 2009.

\bibitem{boden2004creative}
Margaret~A Boden.
\newblock {\em The creative mind: Myths and mechanisms}.
\newblock Routledge, 2004.

\bibitem{urban2005assessing}
Klaus~K Urban.
\newblock Assessing creativity: The test for creative thinking-drawing production (tct-dp).
\newblock {\em International Education Journal}, 6(2):272--280, 2005.

\bibitem{takaffoli2024generative}
Macy Takaffoli, Sijia Li, and Ville M{\"a}kel{\"a}.
\newblock Generative ai in user experience design and research: how do ux practitioners, teams, and companies use genai in industry?
\newblock In {\em Proceedings of the 2024 ACM Designing Interactive Systems Conference}, pages 1579--1593, 2024.

\bibitem{chromik2020ml}
Michael Chromik, Florian Lachner, and Andreas Butz.
\newblock Ml for ux?-an inventory and predictions on the use of machine learning techniques for ux research.
\newblock In {\em Proceedings of the 11th Nordic Conference on Human-Computer Interaction: Shaping Experiences, Shaping Society}, pages 1--11, 2020.

\bibitem{holmquist2017intelligence}
Lars~Erik Holmquist.
\newblock Intelligence on tap: artificial intelligence as a new design material.
\newblock {\em interactions}, 24(4):28--33, 2017.

\bibitem{bourgault2023exploring}
Sam Bourgault and Jane E.
\newblock Exploring the horizon of computation for creativity, 2023.

\bibitem{lo2023impact}
Chung~Kwan Lo.
\newblock What is the impact of chatgpt on education? a rapid review of the literature.
\newblock {\em Education sciences}, 13(4):410, 2023.

\bibitem{sandhaus2024co}
Hauke Sandhaus, Quiquan Gu, Maria~Teresa Parreira, and Wendy Ju.
\newblock Co-designing with algorithms: Unpacking the complex role of genai in interactive system design education.
\newblock {\em arXiv preprint arXiv:2410.14048}, 2024.

\bibitem{das2020opportunities}
Arun Das and Paul Rad.
\newblock Opportunities and challenges in explainable artificial intelligence (xai): A survey.
\newblock {\em arXiv preprint arXiv:2006.11371}, 2020.

\bibitem{ledo2018evaluation}
David Ledo, Steven Houben, Jo~Vermeulen, Nicolai Marquardt, Lora Oehlberg, and Saul Greenberg.
\newblock Evaluation strategies for hci toolkit research.
\newblock In {\em Proceedings of the 2018 CHI conference on human factors in computing systems}, pages 1--17, 2018.

\end{thebibliography}
\nocite{*}

\end{document}